\documentclass[aps,prl,twocolumn,superscriptaddress]{revtex4}
\usepackage{graphicx}
\usepackage{amsmath}
\usepackage{amssymb}
\usepackage{colordvi}
\usepackage{mathrsfs}
\usepackage{bm}
\usepackage{verbatim}
\usepackage{dcolumn}
\usepackage{bm}
\usepackage{epsfig}
\usepackage{subfigure}

\begin{document}
\title{Perfectly Conducting Graphene Electronic Waveguide with Curved Channels}
\author{Vahid Mosallanejad}
\affiliation{CAS Key Laboratory of Quantum Information, and Synergetic Innovation Center of Quantum Information and Quantum Physics, University of Science and Technology of China,Chinese Academy of Sciences, Hefei 230026, China}
\author{Ke Wang}
\affiliation{ICQD, Hefei National Laboratory for Physical Sciences at Microscale, and Synergetic Innovation Center of Quantum Information and Quantum Physics, University of Science and Technology of China, Hefei, Anhui 230026, China}
\affiliation{CAS Key Laboratory of Strongly-Coupled Quantum Matter Physics and Department of Physics, University of Science and Technology of China, Hefei, Anhui 230026, China}
\author{Zhenhua Qiao}
\email[Correspondence author:~~]{qiao@ustc.edu.cn}
\affiliation{ICQD, Hefei National Laboratory for Physical Sciences at Microscale, and Synergetic Innovation Center of Quantum Information and Quantum Physics, University of Science and Technology of China, Hefei, Anhui 230026, China}
\affiliation{CAS Key Laboratory of Strongly-Coupled Quantum Matter Physics and Department of Physics, University of Science and Technology of China, Hefei, Anhui 230026, China}
\author{Guoping Guo}
\email[Correspondence author:~~]{gpguo@ustc.edu.cn}
\affiliation{CAS Key Laboratory of Quantum Information, and Synergetic Innovation Center of Quantum Information and Quantum Physics, University of Science and Technology of China,Chinese Academy of Sciences, Hefei 230026, China}
\date{\today}

\begin{abstract}
We theoretically investigate the electronic transport properties of curved graphene waveguides by employing non-equilibrium Green's function techniques. We systematically  study the dependence of the confined waveguide modes on the potential difference, the width of waveguide and side barrier. Through two-terminal electronic transport calculations, we show that the conductance of confined waveguide modes is rather robust against the bending degree of waveguide, in consistent with the band insensitivity to the side barrier. This finding of the perfectly conducting channels strongly suggests the possibility of applying the graphene waveguide in the design of low-power nanoelectronics.
\end{abstract}

\maketitle

\textit{Introduction---.} Due to its unique linear Dirac dispersion, graphene exhibits excellent physical properties, including the electronic, mechanical and optical aspects~\cite{novoselov2004electric,novoselov2005two,katsnelson2006chiral,young2009quantum}. All these made graphene a promising candidate for designing future electronics. However, the intrinsic zero-gap of graphene hindered the possibility of replacing silicon to realize the basic on/off function of semiconductors. Subsequent efforts have been made to explore various two-dimensional materials that naturally habour bulk band gaps, e.g., silicene~\cite{Silicene-Review} and transition metal dichalcogenide MX$_2$~\cite{TMDC-Review}. Although a variety of new two-dimensional materials become constantly proposed in theory and discovered in experiment, the irreplaceable nature of extremely high mobility at room temperature~\cite{geim2008carbon} has attracted overwhelming attention from both theoretical and experimental researchers to continue exploiting graphene-based electronics via external manipulation after a dozen years ever since its first experimental discovery.

The intrinsic Dirac-cone feature makes graphene an ideal platform to study topologically nontrivial phases that are suitable for dissipationless electronics. So far, there are several approaches that drive the massless Dirac-cone to be massive in graphene~\cite{Qiao-Review}. For examples, by introducing intrinsic spin-orbit coupling, quantum spin-Hall effect (the ancestor of $\mathbb{Z}_2$ topological insulators~\cite{review-TI}) was proposed~\cite{QSHE-Graphene}; by introducing extrinsic Rashba spin-orbit coupling and time-reversal broken ferromagnetism, quantum anomalous Hall effect was proposed~\cite{QAHE-Graphene}. However, these spin-orbit coupling induced topological phases are not  yet experimentally observed due to the difficulty of inducing relatively large spin-orbit couplings in graphene from proximity effect~\cite{YIG-JingShi,LMO-Zeng}. Another controllable degree of freedom to open band gap in graphene is to introduce inter-valley scattering from periodic doping or lattice-matched substrate~\cite{Gap-Valley}, which is still out of control in the current experimental techniques. The third degree of freedom that opens bulk gap in graphene is the top/bottom layers in multilayer graphene~\cite{YuanboZhang-Bilayer} or AB sublattices in monolayer graphene~\cite{AB-gap}. By spatially varying the applied potential difference in bilayer graphene or AB staggered potentials in monolayer graphene, the topological confinement states (or named as topological zero-line mode) along the zero line of potential difference were theoretically proposed~\cite{zero-line-theory} and recently observed in experiments~\cite{zero-line-experiment}.

\begin{figure}
\includegraphics[width=8.5cm]{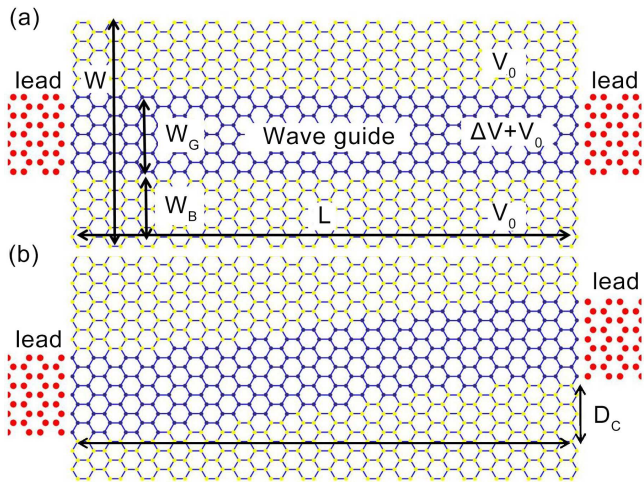}
\caption{Schematic plots of (a) straight and (b) curved waveguides in armchair graphene nanoribbons, being connected with left and right leads extended from the waveguides. Blue atoms label the graphene waveguide. $W$, $W_{\rm G}$ and $W_{\rm B}$ indicate the widths of graphene nanoribbon, waveguide, and side barriers, respectively. $L$ represents the device length. $\Delta V$ is the potential difference between waveguide and barriers, and $V_{\rm 0}$ is the on-site potential in both side barriers. $D_{\rm C}$, the relative shift between left and right terminals, characterizes the degree of bending in curved waveguides.}
\label{Fig1}
\end{figure}

Different from the extreme difficulty in designing structural geometry of topological zero-line modes, graphene waveguide~\cite{hartmann2010smooth, wu2011electronic, myoung2011magnetically} can be easily achieved by externally applying a hard-wall square-shape potential well in transverse direction [see Fig.~\ref{Fig1}(a)]. The electronic transport along the graphene waveguide resembles to the propagation of the electromagnetic waves in an optical media~\cite{dragoman2010polarization}. So far, both theoretical and experimental studies have been mainly concentrated on the straight graphene waveguides~\cite{pereira2006confined, he2015guided,myoung2011magnetically, hartmann2010smooth,zhang2009guided}. And the ballistic transport of the confined states in the quantum well has been experimentally demonstrated~\cite{williams2011gate,rickhaus2015guiding}. It is known that one of the most striking features of optical fibers is that the electromagnetic waves can efficiently transmit in any kinds of bending. This was directly applied in the field of information communication. Analogous to a bending optical fiber~\cite{marcuse1978length}, the curved graphene waveguide may have application prospects in nano-sized integrated systems and electronic interferometry.

In this Letter, we study the electronic transport properties in curved graphene waveguides by employing non-equilibrium Green's function techniques. Graphene waveguides can be readily prepared by simply applying an external electric gate along a finite-sized graphene nanoribbon within the graphene flake. We systematically investigate the dependence of the confined waveguide modes on the potential difference $\Delta V$, the width of waveguide $W_{\rm G}$ and the width of the side barrier $W_{\rm B}$, and find that the electronic bands are insensitive to the variation of the width of side barrier $W_{\rm B}$. By adjusting the external gating, tunable graphene waveguides can be realized. Through a two-terminal electronic transport calculation, we show that the nearly quantized conductance (ballistic transport) is quite robust against the bending degree of waveguide, in consistent with the band insensitivity to the side barrier. Our findings strongly demonstrate the possibility of designing the graphene-waveguide based low-power nanoelectronics.

\begin{figure}
\includegraphics[width=8.5cm]{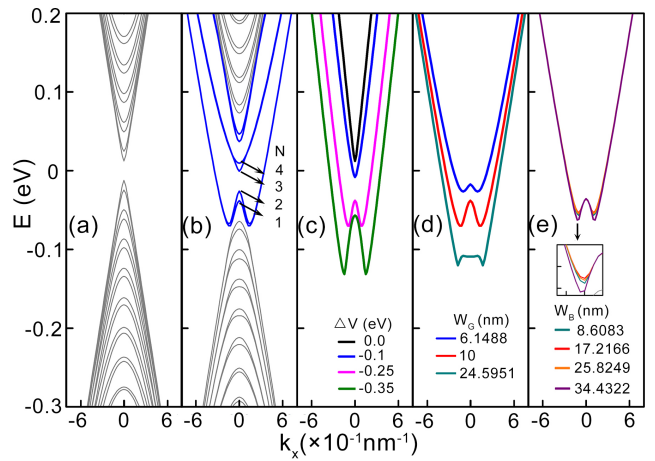}
\caption{(a) Band structure of armchair nanoribbon with a width of $W=50$~nm, with the supercell including 818 atoms. (b) Band structure of armchair nanoribbon harboring a waveguide as displayed in the central region of Fig.~1 (a), where $W_{\rm G}=10$~nm and $\Delta V=-0.25$~eV. The blue bands indicate the modes arising from the waveguide. (c) Relative shift of the first subband due to the variation of $\Delta V$ in the waveguide. (d) First subband mode dispersion for different waveguide thickness $W_{\rm G}$. (e) First subband dispersion for different widths of barriers $W_{\rm B}$. In (c)-(d), the parameters are the same as those in panel (b) except the varying parameter.}
\label{Fig2}
\end{figure}

\textit{System Hamiltonian and Band Structure---.} Band structure can usually provide fundamental information in understanding the electronic transport properties. Without loss of generality, we consider the armchair grephene nanoribbons in our following studies, where valleys $K$ and $K'$ are mixed and indistinguishable. Figures~\ref{Fig1}(a) and \ref{Fig1}(b) respectively display the schematic plots of straight and curved graphene waveguides, where the confined potential $\Delta V$ is applied at the blue colored atomic sites. The $\pi$-orbital tight-binding model Hamiltonian of graphene in the presence of gate voltage can be written as:
\begin{equation}
H=- t \sum_{\langle {ij} \rangle} a_i^{\dagger} a_j + \sum_{i} v_i a_i^{\dagger} a_i ,
\end{equation}
where $a_i^{\dagger}$ and $a_i$ are the creation and annihilation operators at $i$-th atomic site, respectively. $v_i$ is the on-site potential energy, which is set to be zero for reference at the side barrier region and $\Delta V$ in the waveguide region. The nearest-neighbor hopping energy is set to be $t= 2.78$~eV. Due to the relatively large width of the system considered, extra Hamiltonian modification for the edge states or other imperfections is not included\cite{gunlycke2008tight,touski2013substrate}.

\begin{figure}
\includegraphics[width=8.5cm]{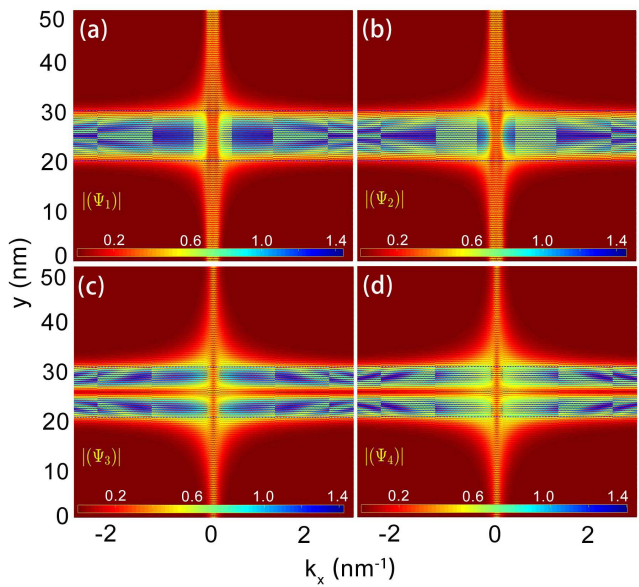}
\caption{Spatial distribution plot of the wavefunctions across the armchair ribbon as a function of $k_x$ for the first four bands as highlighted in Fig.~\ref{Fig2}(b). One can see that all the bands are confined inside the waveguide region except the $k_x$ points  closing to 0. The parameters are respectively $W=50$~nm, $W_{\rm G}=10$~nm, and $\Delta V=-0.25$~eV. Dashed blue lines are inserted to identify the boundaries of the quantum well or waveguide}
\label{Fig3}
\end{figure}

%% 2 (a)
By exactly diagonalizing the tight-binding Hamiltonian of the graphene nanoribbon, the band structure can be obtained. Figure~\ref{Fig2}(a) displays the band structure of armchair ribbon with a well
known conic-shape subbands encapsulated in the celebrated Dirac conicband $E(k_x)=\pm\hbar \nu_F|k_x|$. Here, each unit cell includes 818 atoms, corresponding to a width about $W=50$~nm. By applying a gate voltage of $\Delta V=-0.25$~eV at the region of $W_{\rm G}=10$~nm inside the armchair ribbon, a graphene waveguide is produced and the resulting bands are displayed in Fig.~\ref{Fig2}(b). One can observe that the confined modes (blue colored bands) arise inside the graphene waveguide, and their mixture with the valence bands results in the asymmetry between the conduction and valence bands in the low-energy region. Our further exploration indicates that the confined subbands are insensitive to the ribbon width, beyond the dispersive area with the conduction subbands exhibiting negative slopes.

Now, let us see how the confined waveguide modes evolve as a function of the applied gate voltage $\Delta V$. Figure~\ref{Fig2}(c) displays the first conduction subband variation along with the increase of $\Delta V$. One can see that for relatively small potential, e.g. $\Delta V =-0.1$ eV, the first conduction subband moves downward; for large potentials exceeding the critical point, e.g. $\Delta V = -0.25$~eV and -0.35~eV, the first conduction band becomes overlapped with the valence bands, leading to a band inversion. Figure~\ref{Fig2}(d) displays the evolution of the first conduction band for different waveguide widths $W_{\rm G}=$6.1, 10.0, and 24.6~nm at fixed $\Delta V=-0.25$~eV. One can find that for narrow waveguides, the band inversion occurs at the position closing to the charge neutrality point; while for wide waveguide, the band inversion occurs at the energy further away from the charge neutrality point, therefore giving rise to a wide energy interval for ballistic transport. Figure~\ref{Fig2}(e) displays the dependence of the first conduction band on the width of the side barrier $W_{\rm B}$. One can clearly see that the confined waveguide modes are insensitive against the variation or asymmetry of the width of side barrier $W_{\rm B}$ (see Inset of panel (e) for magnification of small variation), indicating the possible existence of perfectly conducting (i.e. ballistic transport) waveguide channels in randomly formed graphene waveguides.

\begin{figure}
\includegraphics[width=8.5cm]{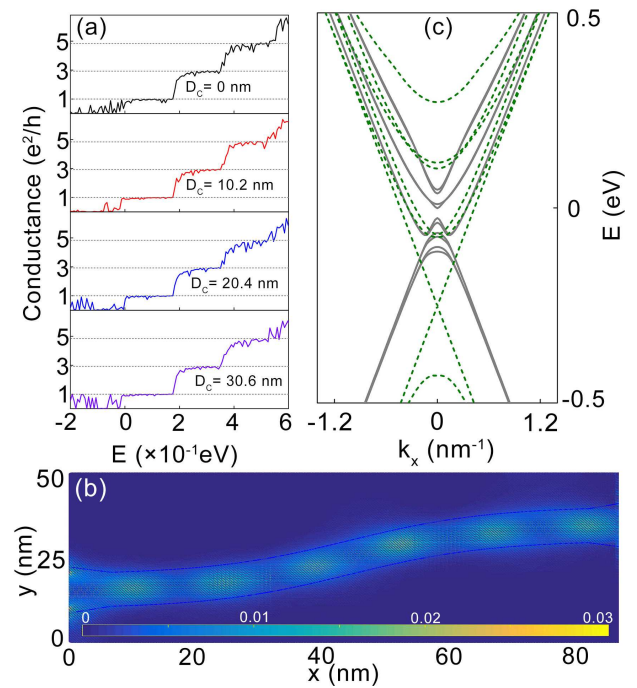}
\caption{(a) Two-terminal conductance as a function of Fermi level for different curved wave guides, i.e., $D_{\rm C}=0$, 10.2, 20.4, and 30.6~nm. The system parameters are chosen to be $L=80$~nm , $W=50$~nm, $W_{\rm G}=10$~nm, and $\Delta V=-0.25$~eV. (b) Spatial distribution of local density of states of the single-mode waveguide. (c) Band matching between the leads (dashed lines) and waveguide area (solid lines) for systems shown in panel (a). Additional on-site potentials of -0.22~eV are added to the leads to align the similar conduction subbands.}
\label{Fig4}
\end{figure}

To show how the waveguide modes are distributed in the potential well, in Fig.~\ref{Fig3} we plot the spatial wavefunction distribution across the armchair ribbon at different wavevectors for the four subbands labelled in Fig.~\ref{Fig2}(b). For clear comparison, two horizontal dashed blue lines are inserted to identify the boundaries of the quantum well or waveguide. Surprisingly, one can clearly see that most of the confined waveguide modes for wave vectors away from $k_x=0$ are perfectly confined inside the waveguide; while the remaining parts are not localized inside the waveguide but distributed in the whole range of the ribbon. This is distinct from the topologically protected edge modes that are completely localized at the system boundaries therefore robust against various external disorders.

\textit{Electronic Transport in Curved Waveguide---.} To explore the robustness of the confined waveguide modes against the bending degree of the graphene waveguide, we use a two-terminal mesoscopic setup to study the conductance as a function of the Fermi level $E_{\rm F}$ for different bending degrees $D_{\rm C}$. The two terminal conductance ${\rm G}(E)$ is expressed as
\begin{equation}
{\rm G}=\frac{e^2}{h} {\rm Tr}[{\Gamma_{\rm L}} {G^{\rm r}} {\Gamma_{\rm R}} {G ^{\rm a}}],
\end{equation}
where $\Gamma_{\rm L,R}$ are the line-width functions coupling the left and right terminals with the central scattering region. The open boundary condition of the left and right terminals is implemented via $\Gamma_{\rm L,R}=i(\Sigma_{\rm L,R}^{\rm r} -\Sigma_{\rm L,R}^{\rm a}) $. The retarded Green function $G^{\rm r}$ is defined as ${G^{\rm r}}=[(E+i\eta)-H-\Sigma_{\rm L} -\Sigma_{\rm R} ]^{-1}$ with $\Sigma_{\rm L,R}$ being respectively self-energies of the semi-infinite left and right terminals~\cite{datta2005quantum}. To effectively examine the robustness of the confined waveguide modes of the curved graphene waveguide, one has to ensure that the terminals can better match the central scattering region to avoid the strong back-scattering. In our consideration, both left and right terminals are exactly extended from the waveguide region as displayed in Fig.~\ref{Fig1}(b). The length and width of the graphene ribbon system are respectively set to be $L=80$~nm and $W=50$~nm. The waveguide width is chosen to be $W_{\rm G}=10$~nm and the applied potential is $\Delta V=-0.25$~eV.

Figure~\ref{Fig4}(a) displays the two-terminal conductance $\rm G$ as a function of the Fermi level for curved waveguide with different bending degrees $D_{\rm C}$=0, 10.2, 20.4 and 30.6~nm. One can see that for the straight graphene waveguide at $D_{\rm C}=0.0$~nm, for $E>0$ the resulting conductance can be approximately quantized to ${\rm G}=1,~3,~5~{e^2/h}$; while for $E<0$ the conductance becomes oscillating due to the strong interference between incoming and backscattering modes and is always less than ${\rm G}=1~{e^2/h}$. It is noteworthy that independent site potentials of $V_L=-0.22$~eV and $V_R=0$~eV are applied to the left and right terminals to match the waveguide modes. Here, the odd quantized conductance plateau is determined by the incoming modes from the terminals [See dashed lines in Fig.~\ref{Fig4}(c)]. To our surprise, when the graphene waveguide becomes gradually curved [see the lower three panels of Fig.~4(a)], negligible backscattering is induced to break the quantization of conductance. On the contrary, the plateau of ${\rm G}=1~e^2/h$ becomes even quantized for a more bending waveguide, i.e. $D_{\rm C}=30.6$~nm. This striking transport phenomenon reflects exactly the zero bend resistance, which may be attributed to the immunity of the confined waveguides to the variation of the width of side barriers and can pose immediate implications in graphene based low-power electronics. 

In Fig.~\ref{Fig4}(b), as an example, we plot the local density of states of the perfectly conducting confined waveguide mode of ${\rm G} \sim1~e^2/h$ at fixed $E_{\rm F}=0.03~eV$ and $D_{\rm C}=20.4~nm$. One can clearly see that the confined waveguide mode only propagates locally inside the waveguide. Figure~\ref{Fig4} (c) shows the band matching between the terminals and central scattering region for Fig.~\ref{Fig4}(a). As mentioned,  additional site potentials of -0.22~eV and 0~eV are respectively added to the terminals' Hamiltonian to align to similar conduction subbands. For different system parameters, the combination of $V_L$ and $V_R$ should be carefully adopted to achieve the zero bend resistance of the graphene waveguide. With the increase of $E_{\rm F}$, the number of bands also increases. The resulting local density of states is no longer perfectly propagating as that in Fig.~\ref{Fig4}(b), but accompanying with very dispersive distribution over the whole scattering area. However, the restriction of the width of the incoming and outgoing terminals ensures the full backscattering of the modes spreading in the whole region and only allows the matched propagating modes, giving rise to the different ballistic transport plateaus. Details about discussing the band matching between terminals and central scattering region are introduced in Supplementary Materials~\cite{SI}.

\textit{Summary---.} To conclude, in this Letter we theoretically investigated the electronic transport properties of the confined waveguide modes in curved graphene waveguide. We first systematically study the dependence of the confined waveguide mode on different system geometric parameters and applied gate potential. We find that the the width of the side barriers does not affect the bands of the confined waveguide modes, which is a hallmark for the possible ballistic transport in a randomly formed graphene waveguide. Further transport calculation gives direct evidence of nearly quantized conductance in curved graphene waveguide. Such a striking zero bend resistance property makes graphene waveguide an ideal platform to design dissipationless or low-power electronic devices.

\textit{Acknowledgments---.} This work was financially supported by the National Key Research and Development Program (Grant No. 2016YFA0301700), NNSFC (Grant Nos. 11625419 and 11474265), and the China Government Youth 1000-Plan Talent Program. The Supercomputing Center of USTC is gratefully acknowledged for high-performance computing assistance. This was also supported by Chinese Academy of Sciences and The World Academy of Science for the advancement of science in developing countries.

V. M. and K. W. contributed equally to this work.

\end{document}